\documentclass[aps,showpacs,twocolumn,superscriptaddress]{revtex4}
\usepackage{dcolumn}   
\usepackage{graphicx}
\usepackage{epsfig}
\usepackage{amsfonts}
\usepackage{amsmath,amssymb}
\usepackage{bm}
\usepackage{longtable}

\newcommand{\ds}{\displaystyle}
\newcommand{\rvec}{\bm{r}}

\begin{document}
\title{Natural Units For Nuclear Energy Density Functional Theory} 

\author{M.~Kortelainen}
\affiliation{Department of Physics \& Astronomy, University of Tennessee, 
Knoxville, Tennessee 37996, USA}
\affiliation{Physics Division,  Oak Ridge National Laboratory, Oak Ridge, 
Tennessee 37831, USA}

\author{R.J.~Furnstahl}
\affiliation{Department of Physics, Ohio State University, Columbus, Ohio 43210}

\author{W.~Nazarewicz}
\affiliation{Department of Physics \& Astronomy, University of Tennessee, 
Knoxville, Tennessee 37996, USA}
\affiliation{Physics Division,  Oak Ridge National Laboratory, Oak Ridge, 
Tennessee 37831, USA}
\affiliation{Institute of Theoretical Physics, Warsaw University, PL-00681, 
Warsaw, Poland}

\author{M.V.~Stoitsov}
\affiliation{Department of Physics \& Astronomy, University of Tennessee, 
Knoxville, Tennessee 37996, USA}
\affiliation{Physics Division,  Oak Ridge National Laboratory, Oak Ridge, 
Tennessee 37831, USA}

\date{\today}

\begin{abstract}
Naive dimensional analysis based on chiral effective theory, when adapted
to nuclear energy density functionals, prescribes natural units 
and a hierarchy of contributions that could be used to
constrain fits of generalized functionals. 
By applying these units,
a large sample of Skyrme parametrizations is examined
for naturalness, which is signaled by dimensionless
coupling constants of order one.
The bulk of the parameters are found to be natural, with
an underlying scale consistent with other determinations.
Significant deviations from unity are associated with
deficiencies in the corresponding terms of particular functionals or with an
incomplete optimization procedure.
\end{abstract}


\maketitle

{\it Introduction.}
New experimental data for atomic nuclei throughout the nuclear mass
chart are becoming available thanks to radioactive nuclear
beam efforts worldwide. This is coupled with increasingly sophisticated 
theoretical descriptions of low-energy nuclear phenomena \cite{bluebook,scidac}.
These developments drive higher requirements for the quality
and predictive power of nuclear structure investigations.

Nuclear density functional theory (DFT) \cite{Bender03} is the only 
available theoretical tool for the microscopic description of nuclear 
properties that spans the full nuclear mass chart. The non-relativistic 
Skyrme energy density functional (EDF) is based on local nuclear densities
and currents and is specified by a set of coupling constants.
There are many sets of Skyrme parameters determined through
different optimizations to experimental energies, radii, and other
nuclear observables. After more than twenty years of
experience, the standard Skyrme EDF has proved to be fairly successful in the
overall description of  experimental data. At the same time,
its limitations  have become well established, and the quest for
better accuracy and stable predictive power motivates  going beyond
the standard Skyrme functional~\cite{Carlsson:2008gm,stoitsov-scidac}.

New developments in nuclear DFT are inevitably associated with an
 optimization of EDF parameters to a selected set
of experimental data. 
This is problematic if more general density dependencies and
higher powers of gradients lead to an explosion of 
new parameters without control over their relative importance.
A possible solution is to organize generalizations of the Skyrme functional
by effective field theory principles that exploit the separation
of scales and so establish a hierarchy of contributions.
One such approach for low-energy quantum chromodynamics is naive
dimensional analysis (NDA) 
for chiral effective field theory~\cite{FRIAR96b}, which has
been adapted to relativistic nuclear EDFs 
with encouraging results~\cite{FRIAR96,RUSNAK97,Furnstahl:2001un,
Burvenich:2001rh}.

In the NDA approach, a scaling to ``natural units'' is applied
to the functional, which if successful results in
dimensionless parameters of order unity.
This has practical benefits, because in numerical optimization
it is advantageous to have all the parameters close to unity.
But the use of natural units also provides 
guidance on maintaining a hierarchy and preventing
fine tuning where higher orders play off against lower orders, particularly
when linear combinations of the parameters are underdetermined.

The relevance of naturalness for Skyrme functionals was suggested
long ago in Ref.~\cite{Fur97}, but the
resulting natural units have not been widely employed (or validated)
by DFT practitioners.
The first test was very limited in the range of
functionals and focused on isoscalar parameters~\cite{Fur97}.
The goal of this paper is to investigate whether 
natural units apply more generally to existing Skyrme parameterizations, 
including the modern ones, and thereby motivate their use in future 
nuclear DFT developments. As part of this study, an online converter to natural
units has been created at \url{http://massexplorer.org},  where one can
browse and convert the Skyrme forces considered in this paper as well as
try new sets of parameters to check whether they are natural.

{\it Skyrme Energy Density Functional.}
The standard Skyrme energy density can be written in isospin
representation as a sum of kinetic and
potential isoscalar ($t=0$) and isovector ($t=1$) energy density terms
\begin{equation}
     {\cal H}(\rvec) =  \ds\frac{\hbar^2}{2m}\tau_0
       + {\cal H}_0(\rvec)+ {\cal H}_1(\rvec) \;,
     \label{CE0}
\end{equation}
where the time-even part is
\begin{eqnarray}
   {\cal H}_t(\rvec)
    &=& \ds    \left(C_{t0}^{\rho} + C_{t{\rm D}}^{\rho}~ \rho_0^\gamma\right) 
       \rho_t^2 + C_t^{\Delta\rho}\rho_t\Delta\rho_t
      +C_t^{\tau} \rho_t\tau_t
    \nonumber \\[5pt]
    & & \null + \ds  \tfrac{1}{2} C_t^{   J}            \bm{J}_t^2
    +        C_t^{\nabla J}        \rho_t\bm{\nabla}\cdot\bm{J}_t \;,
   \label{CE}
\end{eqnarray}
The isospin index $t=\{0,1\}$ labels isoscalar and isovector densities
$\rho_t$, $\tau_t$, and $\bm{J}_t$,  respectively. The standard
definitions of these local densities can be found in Ref.~\cite{Perlinska04}. 
The energy density of  Eq.~(\ref{CE0}) depends on 13 parameters; that is, 12 
coupling
constants and one exponent $\gamma$,
\begin{equation}\displaystyle
  \{C_{t0}^{\rho}, C_{t{\rm D}}^{\rho}, C_t^{\Delta\rho}, 
  C_t^{\tau},C_t^{   J}, C_t^{\nabla J},   \gamma\} \;,
  \label{C-parameters}
\end{equation}
which are typically obtained by adjusting the functional to produce certain
properties in finite nuclei and/or in infinite nuclear matter. Historically,
the Skyrme force and the Skyrme functional derived from it
were defined by using the $\{t_{n},x_{n}\}$ parametrization. The link
between these two representations can be found in Ref.~\cite{Bender03}.
The most general Skyrme functional also contains a time-odd part 
with associated time-odd
coupling constants, which become relevant in nuclear states
with nonzero angular momentum (e.g., odd-mass nuclei)  \cite{Bender03}. 
In this work we consider only the time-even part and leave 
the time-odd coupling constants as a subject of future study.

{\it Natural units.}
Following Ref.~\cite{Fur97}, we
scale the Skyrme coupling constants by
analogy to an effective  Lagrangian where each term is schematically
written as
\begin{equation}
   g\left[\frac{\psi^{\dag}\psi}{\Lambda f_{\pi}^{2}}\right]^{l} 
   \left[\frac{\nabla}{\Lambda}\right]^{n}
   \Lambda^{2}f_{\pi}^{2} \;,
   \label{eq:NDA}
\end{equation}
with $g$ being a dimensionless coupling constant and $f_\pi \approx
93$ MeV is the pion decay constant. 
The momentum $\Lambda$ characterizes the
breakdown scale of the chiral effective theory.
As such, it is expected to be in the range 
$500\,\mbox{MeV} < \Lambda < 1000\,\mbox{MeV}$.
Naturalness implies that $g$ should be of order unity, 
which in practice roughly means between 1/3 and 3 (unless there
is a symmetry reason making $g$ small).
If natural, Eq.~(\ref{eq:NDA}) implies a hierarchy of terms
with a density expansion (powers of $l$) and a 
gradient expansion 
(powers of $n$)~\cite{FRIAR96,RUSNAK97,Furnstahl:2001un,
Burvenich:2001rh,Fur97}.

The conversion of the Skyrme couplings to natural units
is accomplished in the present work
by multiplying each by a scaling factor
\begin{equation}
  S = f_{\pi}^{2(l-1)}  \Lambda^{n+l-2} \;,
  \label{nu}
\end{equation}
where $l$ is the power of densities in the corresponding
term and $n$ is the number of derivatives for that term. In
Ref.~\cite{Fur97} only functionals with integer powers of the 
density-dependent term were considered. Here we generalize the scaling
to include also  fractional  powers $\gamma$ used in Skyrme functionals
by setting $l=2+\gamma$ for the density-dependent term. 
At present this is just a prescription.  
In studies of dilute fermion systems in a harmonic trap, it was
shown that terms with fractional powers in a perturbative functional followed
scaling rules~\cite{Bhattacharyya:2004qm}, 
but this has not yet been derived for the nuclear case.

We also
generalize our analysis to the isovector coupling constants,
which highlights the issue of possible additional numerical
factors in the NDA prescription of
Eqs.~(\ref{eq:NDA}) and (\ref{nu}).
A direct extension to the isovector channel scales the isovector
coupling constants with the same scale factor $S$ as the corresponding
isoscalar couplings.
However, in past applications of the NDA to relativistic meson
and point coupling models, the isovector prescription included an additional
factor of four.
This arises from the construction                 
of the Noether current for an isospin transformation, which in                 
the conventional normalization has a 1/2                    
with each $\tau$ matrix (so that $[T_a,T_b] = i\epsilon_{abc}T_c$              
implies $T_a = \tau_a/2)$.  Then the isovector current is                        
$1/2 \overline\psi\gamma^\mu\tau\psi$ and so $1/2(\rho_p - \rho_n)$ is          
the corresponding charge density used in the naturalness               
analysis.                                       
Although this may be no more than a theoretical prejudice, the empirical
observation in other NDA tests was that the scaled constants
consistently came out closer to unity~\cite{Furnstahl:1999rm}.
In the present study we consider both isovector scalings.

More generally, to decide on possible additional numerical
factors we rely on the correspondence of Skyrme EDF terms
to those from a non-relativistic reduction of a relativistic formulation
(e.g., meson exchange with masses of order $\Lambda$).
For example, one might wonder if the spin-matrix $\sigma$ should lead
to extra scaling factors between scalar terms and those
involving the vector densities ${\bm J}_{t}$. 
We find that such terms arise with the
same relative factor as terms with $\rho\tau$ and so we scale
them the same.

For all coupling constants entering the standard functional of
Eqs.~(\ref{CE0}) and (\ref{CE}), 
one has $l=2$
except for the density-dependent constant $C_{t{\rm D}}^{\rho}$, for which
$l=2+\gamma$. 
Similarly, the power is $n=0$ for $C_{t}^{\rho}$, while  for all
other constants $n=2$. In this way, scaling all coupling constants 
$C^\sigma_t$  with the associated factors $S^\sigma$, 
$\sigma=\{\rho,\Delta\rho,\tau,\nabla J\}$ yields dimensionless constants
$S^\sigma C^\sigma_t$. 
The small ranges for $l$ and $n$ preclude testing the fine details of
the NDA scaling hypothesis.
However, by making a global analysis of Skyrme parametrizations,
we can check for consistency, for trends and exceptions to naturalness, 
and for a preferred range of $\Lambda$.
When parameter sets for extended functionals that include higher-order
derivatives \cite{Carlsson:2008gm} and/or higher powers of density \cite{Biruk}  are available, more 
definitive tests of natural scaling will be possible.

The list of 
functionals considered is given in
Table~\ref{t:skyrmelist}.  In this table we also categorize
the functionals based on the strategy used to determine
the couplings. 

\begin{table}[ht]
\begin{center}
\caption{List of Skyrme functionals and categories used in this study.
The categories are:
a) masses of double-magic nuclei 
(includes $^{90}$Zr, $^{116}$Sn, $^{124}$Sn, and $^{140}$Ce) used in the fit;
b) masses of non-double-magic nuclei used in the fit;
c) charge radii used in the fit;
d) single-particle energies used in the fit;
e) symmetric infinite nuclear matter constrains considered in the fit;
f) asymmetric infinite nuclear matter constrains considered in the fit;
g) surface properties (neutron skin, fission barriers, etc.) considered in the 
fit;
h) pairing was present in the fit;
i) some parameters were fixed in the fit;
j) parameters extrapolated or fine-tuned from an existing force or functional.}
\label{t:skyrmelist}
\begin{tabular}{llrr}
\hline
Index & Functionals & Categories & Ref\\
\hline
 1--2   & SkT3, SkT6     &  a d i j         & \cite{SKT1-9} \\
 3      & SkM            &  a c e f g i     & \cite{SKM} \\
 4      & SkM*           &  g j             & \cite{SKMS} \\
 5--6   & SGI, SGII      &  d e j           & \cite{SGI-II} \\
 7      & HFB9           &  a b f h i       & \cite{HFB9} \\ 
 8--9   & SI, SII        &  a c d e f i     & \cite{SI-II} \\
 10     & SkA            &  a c d e g i j   & \cite{SKA} \\
 11     & HFB16          &  a b c f h i     & \cite{HFB16} \\
 12     & SkT            &  a b d e g h i   & \cite{SKT} \\
 13--16 & SLy4--7        &  a c d e f i     & \cite{SLY4-7} \\
 17--18 & SkI1--2        &  a b c d f g i   & \cite{SKI1-5} \\
 19--20 & SkI3--4        &  a b c d f g i   & \cite{SKI1-5} \\
 21     & SkI5           &  a b c d f g i   & \cite{SKI1-5} \\
 22--27 & MSk1--6        &  a b f h i       & \cite{MSK1-6} \\
 28--29 & SIII, SIV      &  a c i           & \cite{SIII-VI} \\
 30--31 & SV, SVI        &  a c i j         & \cite{SIII-VI} \\
 32--33 & SLy230a,b      &  a c d e f i     & \cite{SLY230A-B} \\
 34--39 & E, E$_{\sigma}$, Z, Z$_{\sigma}$, R$_{\sigma}$, 
          G$_{\sigma}$   &  a c d g i       & \cite{E-GSIGMA} \\
 40     & SkP            &  a b c e f h i   & \cite{SKP} \\
 41--42 & SkO,SkO'       &  a b c d f g i   & \cite{SKO-P} \\
 43     & SV-min         & a b c d g h      & \cite{SVMIN} \\      
 44     & SkO$_{T''}$    &  i j             & \cite{SKOTPP} \\
 45     & SkMP           &  a j             & \cite{SKMP} \\
 46--47 & SkX, SkX$_{\rm c}$ &  a b c d e f & \cite{SKX-C} \\
 48     & RATP           &  a d e f i       & \cite{RATP} \\
\hline
\end{tabular}
\end{center}
\end{table}

The test of whether we truly have natural units
is whether  $S^\sigma$  makes the values of all scaled constants
$S^\sigma C^\sigma_t$ of order unity. 
Their numerical  values will obviously depend on the
value of the cut-off parameter $\Lambda$~\cite{Fur97}.  
In our global study, the naturalness
criterion can itself be used to extract
the value of $\Lambda$ by minimizing the deviation of the coupling 
constants from unity. 
We consider a logarithmic root-mean-square deviation (RMSD)
\begin{equation}\label{RMSD}
{\rm RMSD}=\sqrt{\frac{1}{N}\sum_{i,\sigma,t}\log_{10}^{2}|C^\sigma_t(i)|},
\end{equation}
because naturalness implies couplings should not be too small as
well as not too large.
If a particular coupling
constant is zero, it is excluded from the logarithmic RMSD.

\begin{figure}[htb]
\includegraphics[width=1\columnwidth]{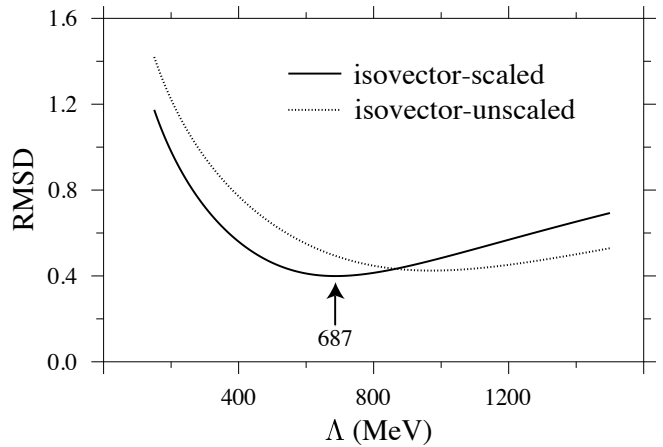}
\caption{Logarithmic RMSD as a function of $\Lambda$ with (scaled) and
 without (unscaled) an extra factor of four for isovector terms. See text for 
details.}
\label{f:rms}
\end{figure}

In Fig. \ref{f:rms} we plot RMSD for 48 EDFs as
a function of $\Lambda$ with (scaled) and without (unscaled)
the extra factor of four for isovector terms.
It can be seen that the two different scalings produce
different optimal $\Lambda$ with the scaled result yielding
a clearer minimum that is numerically more consistent with
studies of relativistic functionals.  However, the minima in the RMSD curves
are quite shallow, so $\Lambda$ cannot be considered to
be sharply determined for the present Skyrme functionals.

\begin{figure*}
\includegraphics[width=2\columnwidth]{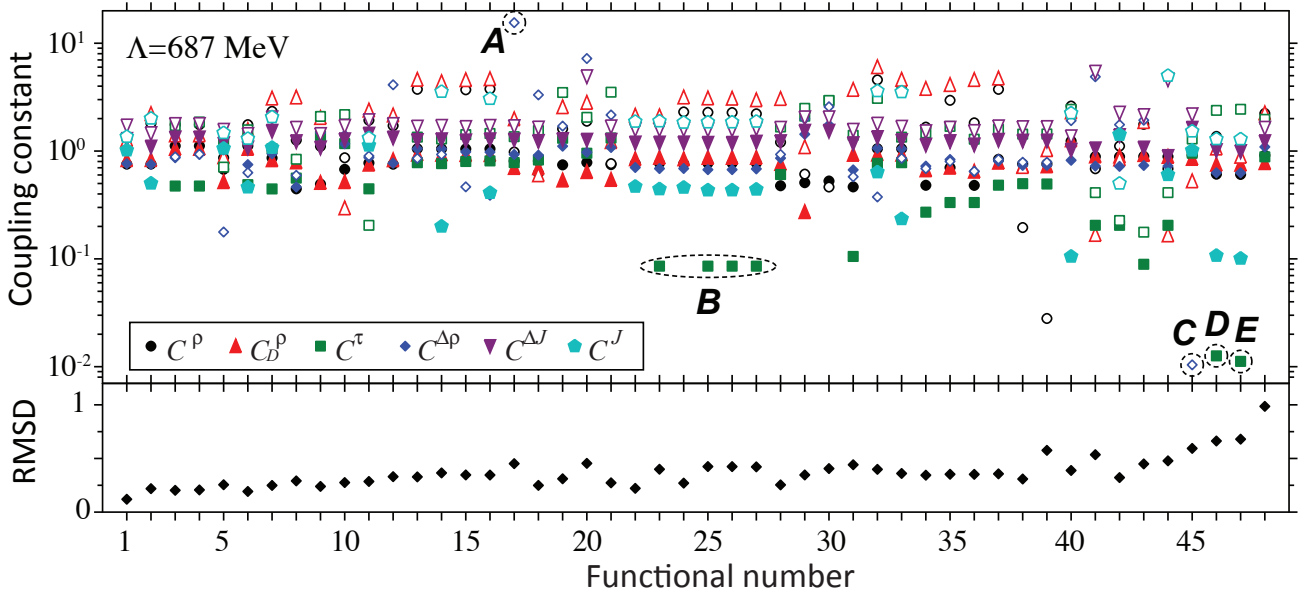}
\caption{(color online) Scaled coupling constants $|C^\sigma_t|$ at 
$\Lambda=$687
MeV (top) and contributions of individual functionals
to the total RMS value (bottom). The filled
symbols refer to the isoscalar coupling constants and empty symbols to
the  isovector ones. The ordering of functionals by index is the same as in
Table~\ref{t:skyrmelist}.}
\label{fig2}
\end{figure*}

In the present study, we choose to use the
scaled isovector coupling constants for which the optimum is
$\Lambda=687$\,MeV (but the precise value does not
affect our conclusions). In Fig.~\ref{fig2} we have plotted the  scaled
coupling constants for all the functionals of Table~\ref{t:skyrmelist}. 
Also, we plot the square-roots  of individual RMSD
contributions given by the functional to the total RMSD value.
It can be seen that the Skyrme
functionals have almost all of their parameter values in the interval 
$(1/3,3)$ with the bulk between $1/2$ and 2.  Exceptions are discussed
below.

We also make a comparison between different representations of two
particular functionals: SIII and HFB16. The parameters of these functionals are
listed in Table~\ref{t:forcecomp} first by using the $(t,x)$-parametrization
and then by the natural units parametrization, obtained from the corresponding
coupling constants. As can be seen, in the
$(t,x)$-parametrization these two functionals seem to be quite different from
each other. However, when expressed  in natural units 
the coupling constants of SIII and HFB16 are order unity.
In Table~\ref{t:forcecomp} we also list in natural units
the average, minimum, and maximum value for each coupling constant
found in the set of 48 functionals. This information
may provide useful insights into the expected values and ranges of coupling
constants for future attempts to fit new functionals.

\begin{table*}[ht]
\begin{center}
\caption{Comparison between the SIII and HFB16 functionals 
in the $(t,x)$-parametrization and using
natural units. We also list for each coupling constant the average, 
minimum, and maximum
value found in the set of 48 functionals.
In all cases, $\Lambda= 687\,$MeV was used.}
\label{t:forcecomp}
\begin{tabular}{lrr|lrr|rrr}
\hline
\multicolumn{3}{c}{$(t,x)$-parameters} & \multicolumn{6}{c}{Couplings in natural 
units} \\
 & SIII & HFB16 & & SIII & HFB16 & Average & Min. & Max. \\
\hline
$t_{0}$ & $-$1128.75 & $-$1837.23       &  $C^{\rho}_{00}$       & $-$0.4767 & 
$-$0.7759 & $-$0.7977 & $-$1.2380 & $-$0.4465 \\
$t_{1}$ & 395.0    & 383.521        &  $C^{\rho}_{10}$       &  1.2076 &  1.9295 
&  1.7656 & $-$0.9795 &  4.5761 \\
$t_{2}$ & $-$95.0    & $-$3.41736       &  $C^{\rho}_{0{\rm D}}$ &  0.7623 &  
0.7509 &  0.7824 &  0.2723 &  1.2616 \\
$t_{3}$ & 14000.0  & 11523.0        &  $C^{\rho}_{1{\rm D}}$ & $-$3.0493 & $-
$2.3825 & $-$2.2010 & $-$6.0116 &  1.9790 \\
$x_{0}$ & 0.45     & 0.432600       &  $C^{\tau}_{0}$        &  0.6059 &  0.4464 
&  0.5606 & $-$0.0856 &  2.9421 \\
$x_{1}$ & 0        & $-$0.824106      &  $C^{\tau}_{1}$        & $-$1.6726 & $-
$0.2048 & $-$0.3626 & $-$2.9469 &  3.5160 \\
$x_{2}$ & 0        & 44.6520        &  $C^{\Delta\rho}_{0}$  & $-$0.8597 & $-
$0.8702 & $-$0.8765 & $-$1.7491 & $-$0.4636 \\
$x_{3}$ & 0        & 0.689797       &  $C^{\Delta\rho}_{1}$  &  0.9301 & $-
$0.8998 & $-$0.5531 & $-$15.5202 & 2.5762 \\
$W_{0}$ & 120.0    & 141.100        &  $C^{\nabla J}_{0}$    & $-$1.2288 & $-
$1.4449 & $-$1.2385 & $-$1.6384 & $-$0.8957 \\
        &          &                &  $C^{\nabla J}_{1}$    & $-$1.6384 & $-
$1.9265 & $-$1.0875 & $-$2.1846 &  5.4272 \\
        &          &                &  $C^{J}_{0}$           &  0      &  1.1300 
&  0.5159 & $-$0.6021 &  1.4212 \\
        &          &                &  $C^{J}_{1}$           &  0      &  1.3208 
&  1.6502 & $-$5.0026 &  3.6049 \\
\hline

\end{tabular}
\end{center}
\end{table*}

{\it Deviations from order unity.}
The deviations of the coupling constants $C^\sigma_t$ from unity
are illustrated in the summary plot in Fig.~\ref{fig2}.
As noted earlier, almost all parameters are found to lie
within the interval of $(1/3,3)$.
In terms of naturalness, we do not observe any significant differences 
between the functionals that are strictly based on the Skyrme force and 
the extended functionals.
However, significant deviations still exist for some 
particular Skyrme functionals
for the coupling constants $C_0^{J}, C_1^{\Delta \rho}$, and
$C_0^{\Delta J}$, and more generally for $C^{\rho}_{1{\rm D}}$,
which appears borderline unnaturally large in many cases. 
If we accept that nuclear functionals are
characterized by naturalness, such deviations could indicate some deficiency in
the associated term of the functional or they could 
simply reflect a specific strategy applied to the optimization procedure. 

While in some cases no definite cause has been identified,
we can identify various examples of unnatural couplings that do have
probable explanations.
For example, it is well known that tensor terms are rather poorly
constrained by the experimental data; most Skyrme functionals
do not include tensor terms at all. The significant deviations seen
in Fig.~\ref{fig2} for the $C_0^{J}$ parameters most likely reflects
this situation.

Another instructive example is the deviation for $C_{1}^{\Delta \rho}$ 
in the  case of SkI1 (case A in Fig.~\ref{fig2}). The optimization 
of SkI1 excluded the isotope shift data while all
other SkI forces (SkI2--SkI5) consider these data.
This isotopic shift data contains the charge radii
difference between $^{40}$Ca and $^{48}$Ca, and $^{208}$Pb and
$^{214}$Pb. All SkI functionals are, however, optimized by using
diffraction radii data, which is closely related to charge radii.
 
The deviation in $C_0^{\tau}$ for SV results from an artificially
imposed vanishing density-dependent term, which results in a too-low 
isoscalar effective mass (0.38). The EDF
RATP demonstrates a clear example of an anomalously small 
$C^{\Delta\rho}_{1}\approx -0.0019$ (not seen on the scale of 
Fig.~\ref{fig2}). This can probably be attributed to the 
fitting procedure, where the focus was mainly on
infinite nuclear matter properties for  astrophysical applications.
Similarly, $C^{\Delta\rho}_{1}$ of SkMP (case C)
is also very small.
This can be linked to the fact that this functional was obtained by
mixing $t$- and $x$-parameters of SkM* and SkP functional and making
small adjustments to the volume part of the functionals. Therefore,
almost no attention was paid to the surface part  either in RATP or
SkMP functionals.

Another example is an anomalously small $C_0^{\tau}$, found in SkX and 
SkX$_{c}$ (cases D and E). This is
due to the fact that in fitting these forces, much emphasis was put on
the single-particle energies. This leads to an effective mass close to
one, and therefore a small coupling constant. Similarly, in the MSk series
(case B) the fit favored effective mass close to one, and
it was therefore set by hand either to 1.0 or 1.05. This, however,
does not imply that single-particle energies are not suitable
observables in the fitting procedure.

Finally, the seemingly unnaturally large $C^{\rho}_{1{\rm D}}$ couplings
may reflect an inadequate treatment of terms with fractional
density dependence.  Alternatively, the fact that the isovector
$C^{\rho}_{10}$ couplings are also sometimes unnatural hints
at a problem with the scaling of terms associated
with pion-range physics.  The density matrix expansion applied
to the leading long-range contributions from chiral effective
field theory may shed light on this issue. 

These examples illustrate that the use of natural units in
nuclear DFT not only introduces  the simplicity
of dimensionless coupling constants and the convenience of their order-unity
values, but also can give  valuable pointers to
potential deficiencies of the physics invoked when constructing and
optimizing the functional.

{\it Conclusions.}
In this study, the coupling constants of a large set of
Skyrme EDFs have been examined for naturalness
as an extension of Ref.~\cite{Fur97}. 
While the limited range of density and gradient terms in 
the standard Skyrme parameterizations means that a definitive
test of NDA chiral naturalness is not possible,
the best functionals are consistent with naturalness and
a scale $\Lambda$ of about 700\,MeV. 
Significant deviations from unity can be associated with
deficiencies in fitting the functional or with specific optimization
procedures. This motivates
using naturalness as a guiding  principle for constructing
new generalized Skyrme functionals.  
An online natural units convertor has been set up at
\url{http://massexplorer.org} as a tool for such applications.
Further investigation is needed to better understand how to treat
non-analytic density dependence and hybrid functionals
where the density matrix expansion is used for long-range
contributions.
Finally, we note that the phenomenologically successful finite-range Gogny
functionals can be accommodated within the Skyrme 
framework~\cite{Dobaczewski:2010qp}, which can be used to
broaden the application of natural units.

%
The UNEDF SciDAC Collaboration is supported by the Office of Nuclear Physics, 
U.S. Department of Energy under Contracts 
Nos.\ DE-FC02-09ER41583 and DE-FC02-09ER41586 (UNEDF SciDAC Collaboration),
DE-FG02-96ER40963 (University of Tennessee),  and by
the National Science Foundation under Grant No.~PHY--0653312.
Computational resources were provided by the National Center for
Computational Sciences at Oak Ridge and the National Energy Research
Scientific Computing Facility.

\end{document}